# Solitons in spiraling Vogel lattices

Yaroslav V. Kartashov,[1,2] Victor A. Vysloukh,[3] and Lluis Torner[1]

[1]*ICFO-Institut de Ciencies Fotoniques, and Universitat Politecnica de Catalunya, 08860 Castelldefels (Barcelona), Spain*
[2]*Institute of Spectroscopy, Russian Academy of Sciences, Troitsk, Moscow Region, 142190, Russia*
[3]*Departamento de Fisica y Matematicas, Universidad de las Americas – Puebla, Santa Catarina Martir, 72820, Puebla, Mexico*

We address light propagation in Vogel optical lattices and show that such lattices support a variety of stable soliton solutions in both self-focusing and self-defocusing media, whose propagation constants belong to domains resembling gaps in the spectrum of a truly periodic lattice. The azimuthally-rich structure of Vogel lattices allows generation of spiraling soliton motion.

The evolution of light beams in materials with a shallow periodic modulation of the refractive index in the direction transverse to the propagation direction, or optical lattices, is well understood [1-3]. This includes linear propagation as well nonlinear self-sustained light states. Solitons emerging from forbidden gaps in the band-gap spectra of many optical lattices have been observed in various waveguide arrays [4-6] and also in reconfigurable optically induced structures [7-10]. The method of optical induction suggested in [7] allows the creation of a variety of refractive index landscapes induced by non-diffracting light patterns that remain invariable along the propagation direction, thus affording a number of new types of stationary soliton families and dynamic soliton phenomena unique to such complex optical lattices [11-19]. Among them are rotary motion of solitons in Bessel photonic lattices [11-14], the formation of strongly elliptical vortices in Mathieu lattices [17], and oscillations of Weber solitons in parabolic lattices [19].

Of particular interest is the propagation of light in strongly distorted asymmetric refractive index landscapes. Such settings, even with unusual symmetries, can be created experimentally by using optical induction [20] or direct waveguide fabrication [21]. Light propagation in such non-periodic refractive index landscapes has been studied in circular, parabolic, and elliptical lattices produced by the corresponding non-diffracting beams and in so-called geometrical potentials appearing due to local deformations of guiding channels [21,22]. Most such refractive index landscapes support well-localized linear modes, a feature that has relevant implications to the existence of solitons.

In this Letter we address the propagation of light in spiraling Vogel optical lattices, whose topology differs remarkably from the shape of lattices considered so far. Single-site excitations in such lattices exhibit strong discrete diffraction acquiring nonzero angular momentum upon evolution, in contrast to discrete diffraction in periodic lattices. Vogel lattices support strongly asymmetric solitons with spiraling tails. Also, solitons are shown to move along the lattice spiral arms.

We consider the propagation of light beams along the $\xi$ axis of a cubic nonlinear medium with an imprinted transverse refractive index modulation that can be described by the nonlinear Schrödinger equation for the dimensionless light field amplitude $q$:

$$i\frac{\partial q}{\partial \xi} = -\frac{1}{2}\left(\frac{\partial^2 q}{\partial \eta^2} + \frac{\partial^2 q}{\partial \zeta^2}\right) + \sigma q |q|^2 - pR(\eta,\zeta)q. \qquad (1)$$

Here $\eta, \zeta$ and $\xi$ are the transverse and longitudinal coordinates normalized to the characteristic beam width and diffraction length, respectively; $\sigma=\pm 1$ corresponds to defocusing/focusing nonlinearity; $p$ is the lattice depth and the function $R(\eta,\zeta)$ describes the refractive index distribution in the lattice. We consider Vogel lattices made of identical super-Gaussian waveguides $R=\sum_{k=0,\infty}\exp\{-[(\eta-\eta_k)^2+(\zeta-\zeta_k)^2]^2/d^4\}$, whose centers $(\eta_k,\zeta_k)$ reside on a parabolic spiral, i.e. $(\eta_k,\zeta_k)=\pm(r_k\cos\phi_k, r_k\sin\phi_k)$, where $r_k=a_1 k^{1/2}$ and $\phi_k=a_2 k$, $k$ is the waveguide ordering number, and $a_{1,2}$ are parameters determining the separation between neighboring spiral arms and the angular separation between neighboring waveguides, respectively. The value $a_2=\pi(3-5^{1/2})$, corresponding to the golden angle, was used by Vogel to describe the structure of a sunflower. An example of a Vogel lattice corresponding to $a_2=\pi(3-5^{1/2})$ is shown in Fig. 1(d); lattices obtained for $a_2=0.06\pi$ are depicted in Figs. 1(e),(f). We consider waveguides with width $d=0.5$ and set $a_1=1.7$ unless otherwise stated.

To understand how a spiraling refractive index distribution affects beam propagation, we first consider the evolution dynamics of narrow, low-power excitations that at $\xi=0$ occupy only the central waveguide of the structure. Figures 1(a)-1(c) illustrate the discrete diffraction that takes place in the lattice, by which light gradually diffracts toward the periphery due to coupling between nearest waveguides. Notice that for the values of the parameter $a_2$ selected in Figs. 1(a)-1(c) the separation between waveguides in the same spiral arm exceeds the separation between waveguides in neighboring arms and thus the most efficient coupling occurs in the direction orthogonal to original spiral. At the same time, by increasing $a_2$ one can realize a situation where coupling occurs only between waveguides from the same arm and light expands only along but not across the spiral - this is the case for the lattice shown in Fig. 1(e). Strong discrete diffraction in the lattice implies the absence of at least well-localized linear modes, since light does not concentrate in the center of the structure. Moreover, the presence of several beams spiraling outward the center of the lattice in the same angular direction that is visible in Fig. 1(c) suggests that the evolution of single-site excitations in Vogel lattice is accompanied by the accumulation of beam orbital angular momentum.

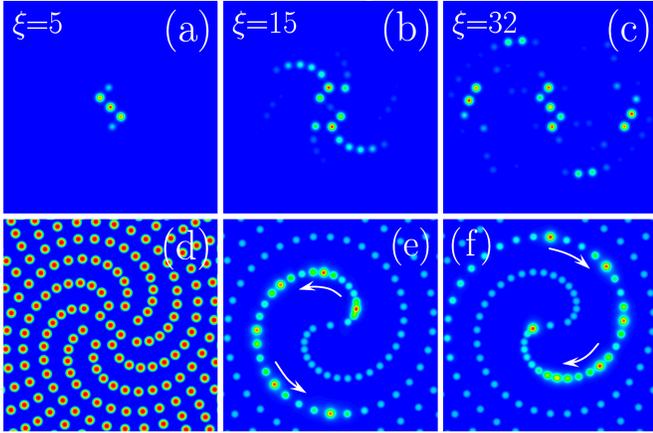

Fig. 1. (a)-(c) Dynamics of propagation of a single-site excitation in a linear Vogel lattice with $p=8$. (d) Example of a Vogel lattice with $a_1=1.7$ and $a_2=\pi(3-5^{1/2})$. Snapshot images showing spiraling soliton motion outward (e) and inward (f) the center of a lattice with $a_1=2.5$, $a_2=0.06\pi$, and $p=3$. In (e) the soliton with $U=2.28$ was set into motion by multiplying the input field with $\exp(i\alpha\zeta)$, with $\alpha=1.4$, while in (f) the soliton with $U=1.49$ was set in motion by means of an input tilt $\exp(i\alpha\eta)$, where $\alpha=0.6$. Snapshot images are superimposed on the lattice shape. White arrows indicate the direction of motion.

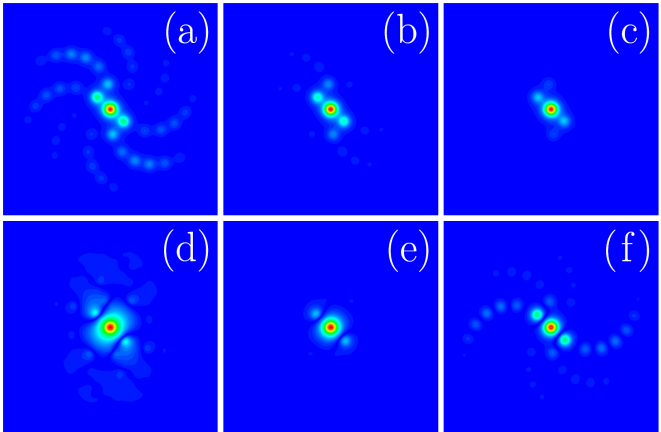

Fig. 2. Field modulus distributions for solitons supported by a Vogel lattice with depth $p=8$ in focusing media with (a) $b=3.01$, (b) $b=3.05$, and (c) $b=3.20$, and defocusing media with (d) $b=-0.30$, (e) $b=0.75$, and (f) $b=1.98$. Solitons from panels (a)-(c) correspond to circles in Fig. 3(a), while solitons from panels (d)-(f) correspond to circles in Fig. 3(b).

Vogel lattices with both focusing and defocusing nonlinearity support various families of steady-state solutions whose field distributions can be written in the form $q=w(\eta,\zeta)\exp(ib\xi)$, where $b$ is the propagation constant. Representative profiles of such solitons residing in the center of the lattice and possessing nonconventional spiraling tails are shown in Fig. 2. In focusing media solitons were found for $b$ values exceeding a certain cutoff $b_0^{\text{low}}$ [see Fig. 3(a) with typical dependence of soliton energy flow $U=\int\int_{-\infty}^{\infty}|q|^2\,d\eta d\zeta$ on $b$]. As $b$ approaches $b_0^{\text{low}}$ solitons strongly expand across the lattice [Fig. 2(a)], while far from cutoff they become well localized [Fig. 2(c)]. The dependence $U(b)$ is nonmonotonic, i.e. there is a threshold for soliton existence, a result that is consistent with the absence of localized linear modes in the center of the lattice. The cutoff $b_0^{\text{low}}$ monotonically grows with increasing $p$ [see lower edge of semi-infinite existence domain shown with white color in Fig. 3(c)], resembling the behavior of stationary states residing in the semi-infinite gap of periodic lattices.

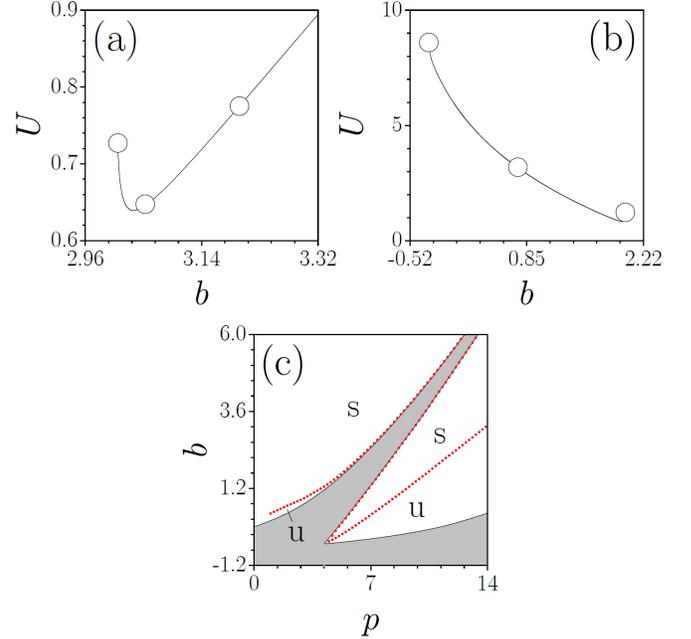

Fig. 3. Energy flow versus propagation constant for solitons in (a) focusing and (b) defocusing media when $p=8$. (c) The domains of soliton existence (white regions), stability (regions marked with "s"), and instability (regions marked with "u") on the plane $(p,b)$. The domains of stability in semi-infinite gap are shown for focusing media, while in the first finite gap they are shown for defocusing media.

In defocusing media we found soliton states only for a limited range of propagation constants $b_1^{\text{upp}}\geq b\geq b_1^{\text{low}}$, above a certain energy flow threshold [Fig. 3(b)]. In contrast to solitons in focusing media, such soliton states exhibit a staggered phase distribution (i.e. neighboring spots in the soliton profile are out-of-phase). Solitons expand considerably across the lattice close to the lower and upper edges of the existence domain [Figs. 2(d) and 2(f)] and remain well localized in its center [Fig. 2(e)]. The spiraling soliton shape becomes most pronounced near the upper edge of the existence domain $b\to b_1^{\text{upp}}$ [Fig. 2(f)]. The domain of solitons existence in defocusing media opens up for sufficiently large lattice depths [see lower white region in Fig. 3(c) located between gray domains where no solitons were found], a result that resembles the behavior encountered in the first finite gap of the eigenvalue spectrum of perfectly periodic two-dimensional lattices. Thus, one concludes that despite the nonconventional spiraling shape of Vogel lattices, the nonlinear excitations supported by them exhibit properties qualitatively similar to those of solitons in periodic lattices.

The stability of the stationary solutions was studied by direct propagation in the presence of small $(\delta q/q\sim 10^{-3})$ input perturbations. The results are summarized in Fig. 3(c) where stability domains are denoted with letters "s" and instability domains are indicated by "u". In focusing media the branch of the $U(b)$ dependence where $dU/db \leq 0$ corresponds to unstable solutions, while the branch where $dU/db > 0$ corresponds to stable solutions. This result is consistent with the fact that the linear eigenvalue problem for perturbations that is obtained from Eq. (1) gives only one purely real eigenvalue leading to an exponential instability in the domain where $dU/db \leq 0$ and does not have complex eigenvalues (which would lead to drift instabilities). The edge of this very narrow instability domain located close to the cutoff $b_0^{\text{low}}$ is indicated in Fig. 3(c) with a red dotted curve. For solitons in defocusing media the structure of instability domains is more complex. There exist a very narrow instability domain close to the upper cutoff $b_1^{\text{upp}}$ [the lower edge of this instability domain is also shown in Fig. 3(c) with a dotted curve nearly coinciding with $b_1^{\text{upp}}$] and a wide instability domain occupying nearly half of the finite gap adjacent to the lower cutoff $b_1^{\text{low}}$. Notice, however, that inside this domain the instabilities may be so weak for $p > 10$ that they do not lead to considerable shape transformations even at $\xi = 10^4$. Solitons from the upper part of the finite gap were found to be stable.

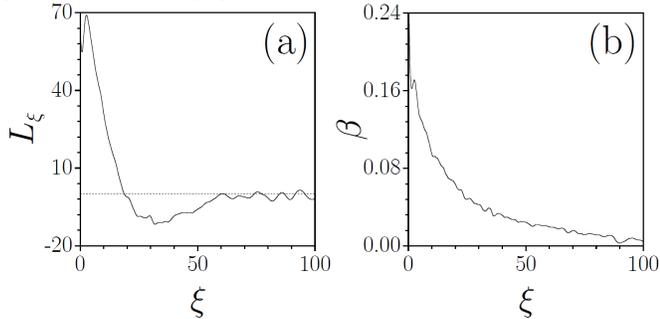

Fig. 4. (a) Orbital angular momentum and (b) angular velocity of spiraling solitons versus $\xi$. All parameters are as in Fig. 1(e). The dashed line in (a) corresponds to $L_\xi = 0$.

The spiraling structure of Vogel lattice allows generation of soliton motion that is not accessible in other lattices. Utilizing the fact that low-energy solitons are relatively mobile in two-dimensional lattices with focusing nonlinearity, we set them into spiraling motion along the lattice by imprinting a suitable phase gradient on them [i.e. by multiplying the input field of a stationary soliton by $\exp(i\alpha\eta + i\gamma\zeta)$ term]. Snapshot images taken at different propagation distances are shown in Figs. 1(e) and 1(f). Due to increasing density of waveguides in the center of the lattice, it is easier to force solitons to spiral towards the center rather than outwards. Spiraling causes radiation, thus solitons gradually slow down. The longitudinal component of the angular momentum acquired by the beams in the lattice, given by the expression $L_\xi = 2\int\int_{-\infty}^{\infty} \text{Im}[q^*(\eta \partial q/\partial\zeta - \zeta \partial q/\partial\eta)]d\eta d\zeta$, evolves according to the equation:

$$\frac{d}{d\xi}L_\xi = -2p\int\int_{-\infty}^{\infty} R(\eta\partial|q|^2/\partial\zeta - \zeta\partial|q|^2/\partial\eta)d\eta d\zeta \quad (2)$$

and gradually vanishes at $\xi \to \infty$ when solitons get trapped in one of the lattice channels [see Fig. 4(a) showing $L_\xi(\xi)$ for soliton spiraling outward lattice center]. This is accompanied by a monotonic decrease of the angular rotation velocity $\beta = d\phi/d\xi$, where $\phi$ is the angular position of soliton center [Fig. 4(b)].

Summarizing, we showed that Vogel optical lattices support stable solitons with nonconventional, nontrivial shapes in both self-focusing and self-defocusing nonlinear media. The very nature of the lattice affords a natural rotational degree of freedom for the solitons, which can be set into motion along curvilinear trajectories by varying the input beam tilt and power. The results obtained here are relevant also for matter waves trapped in spiraling optical lattices.